\documentclass[%
 preprint,
 amsmath,amssymb,
 aps, prl
]{revtex4-1}
\usepackage{graphicx}
\usepackage{dcolumn}
\usepackage{bm}
\usepackage{siunitx}
\usepackage[caption=false]{subfig}

\begin{document}

\preprint{StankiewiczPRLsubmit-2018}

\title{Low Frequency Noise in Randomly Stimulated Asymmetric Oscillators}

\author{Andrzej Stankiewicz}
\email[Electronic mail: ]{Andrzej.Stankiewicz@seagate.com}
\affiliation{Recording Head Group, Seagate Technology\\Bloomington, (Minnesota, USA)}

\date{\today}

\begin{abstract}
A new mechanism of low frequency (1/f-like) noise generation is described and analyzed. It is attributed to higher frequency asymmetric resonance modes, which are stimulated by a random factor, e.g. due to thermal excitation. One-dimensional models of bi-harmonic and non-linear asymmetric oscillators are presented to prove the concept, and a method of detecting the effect is developed. The method is then applied to experimental data, in order to show that the effect exists in magnetoresistive readers. 
\end{abstract}

\maketitle

\paragraph{Introduction -}
Many physical systems and devices are designed to work in a quasistatic regime. This means that the system's adaptation to varying input factors is so fast that the system practically follows its instantaneous equilibrium state. Hence, when these factors remain constant, the system output is expected to be constant, corresponding to the equilibrium state, i.e. a local minimum of the energy. However, at finite temperature, the system actually oscillates around this equilibrium with fluctuating amplitude, generating broadband noise. While these effects are present in any device, historically they did not bring much harm to system signal-to-noise ratio (SNR), as the resonance frequencies were typically much larger than the operational frequency range of the device, and the excitation was very small. As a result, a harmonic oscillator approximation could be applied, which produces thermal noise spectra replicating a damped oscillator resonance curve, with a nearly flat and very weak low-frequency (LF) section \cite{NorrelykkeHarmonicoscillatorheat2011}. This situation has been changing with progress in the development of micro-electro-mechanical systems (MEMS). Miniature devices are much more prone to thermal agitation, making thermal noise a critical factor in their performance \cite{Mohd-YasinNoiseMEMS2010}. 

Noise in electronic devices has been investigated for decades, and its common types (e.g. Johnson, shot or random telegraph) are well understood \cite{KoganElectronicnoisefluctuations1996}. The biggest remaining problem is posed by 1/f (or flicker) noise, which is also common in many devices, but does not have any universal explanation. Theories and models for different systems usually assume that it is a net result of many discrete processes (e.g. charge trapping, carrier scattering, changes in magnetic domain ripples, etc.) that have different relaxation times. Averaging such random processes may lead to a $1/f^{\alpha_H}$ noise profile in low frequency range, where $0 < \alpha_H \lesssim 3$ is a Hooge constant. 

In this Letter, a concept fluctuating asymmetric oscillator as a possible source of 1/f-type noise is developed. While fluctuating non-linear oscillators also have a long research history \cite{DykmanFluctuatingNonlinearOscillators2012}, the attention was usually paid to high frequency properties. It turns out that oscillator asymmetry, combined with random stimulation, inevitably leads to significant LF noise. Let us consider a couple simple examples, with the assumption that output signal is proportional to a generalized displacement $x(t)$, where $t$ denotes time. 

\begin{figure}[!t]
	\centering
	\includegraphics[width=8.6 cm]{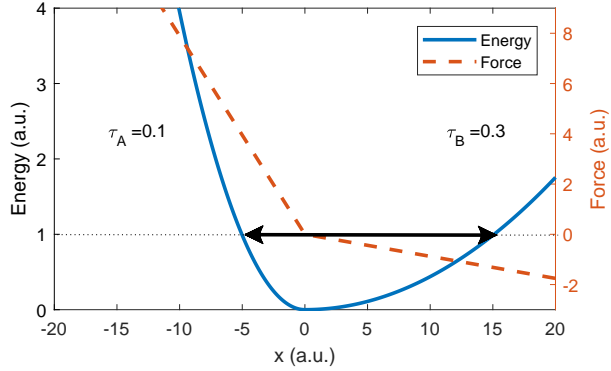}
	\caption{(Color online) The potential energy and returning force for a bi-harmonic oscillator of asymmetry $\lambda=0.5$. Arrows indicate oscillation range, when the system is excited to the energy level of 1.}
	\label{fig:oscEnForce}
	\end{figure}

\paragraph{Bi-harmonic oscillator model -}
We will first consider a bi-harmonic oscillator, which has been previously analysed in \cite{Rucklebiharmonicoscillatorasymmetric2012}. It can be realized in practice as a mass placed (not attached)  between two fully relaxed springs A and B of different stiffness constants. At the equilibrium position ($x=0$) no force is applied on the object. When it is shifted to the left ($x<0$), only the A spring is compressed and the B spring remains in its equilibrium state. For a right shift, only the B spring exerts force (FIG.~\ref{fig:oscEnForce}). Such a setup results in a free oscillation period $\tau_0$ to be composed of two different half-periods $\tau_A/2$ and $\tau_B/2$ corresponding to springs A and B respectively.  

For convenience we introduce an asymmetry parameter $\lambda \in (-1,1)$, such that: 
	\begin{equation}\label{eq:asym}
	\begin{aligned}
    	&\tau_A = \tau_0 (1-\lambda) \quad ,
	\\
     	&\tau_B = \tau_0 (1+\lambda) \quad ,
	\\
    	&\tau_0 =1/f_0=(\tau_A+\tau_B)/2 \quad , 
	\end{aligned}
	\end{equation}
where 
    \begin{equation}\label{eq:lambda}
       \lambda = \frac{\tau_B-\tau_A}{\tau_B+\tau_A} \quad ,
    \end{equation}
and $f_0$ is the oscillator frequency. 
We can therefore set a constant free oscillation period for the bi-harmonic system, and use a single parameter $\lambda$ to control asymmetry. The displacement dependent circular frequency can then be written as:
    \begin{equation}\label{eq:omega}
       \omega_\lambda (x) = 
	   \begin{cases} 
  	      2 \pi f_0/(1-\lambda) & \text{if } x < 0 \\ 
	      2 \pi f_0/(1+\lambda) & \text{if } x \geq 0 
	    \end{cases} \quad . 
    \end{equation}
We also have to define a displacement dependent damping coefficient $\beta(x)$, to be consistent with the fluctuation-dissipation theorem. With $\alpha$ as the damping constant, we obtain: 
	\begin{equation}\label{eq:beta}
	  \beta_\lambda (x) = \alpha \omega_\lambda (x) \quad .
	\end{equation}
The Langevin equation for our model is: 
	\begin{equation}\label{eq:LangevinBi}
	  \ddot{x} + 2 \beta_\lambda (x) \dot{x} + (\omega_\lambda (x))^2 x = \eta (t) \sqrt{\beta_\lambda (x) \omega_\lambda (x)}	\quad ,
	\end{equation} 
where dots are time derivatives, and $\eta (t)$ represents the normalized white noise process. The intensity coefficient of $\eta (t)$ was selected so as to normalize the average excitation energy. The equation can be numerically integrated with one of the standard methods for stochastic differential equations \cite{KloedenNumericalSolutionStochastic2013}. The Euler-Maruyama approximation was applied here.

	\begin{figure}[!t]
	\centering
	\includegraphics[width=8.6 cm]{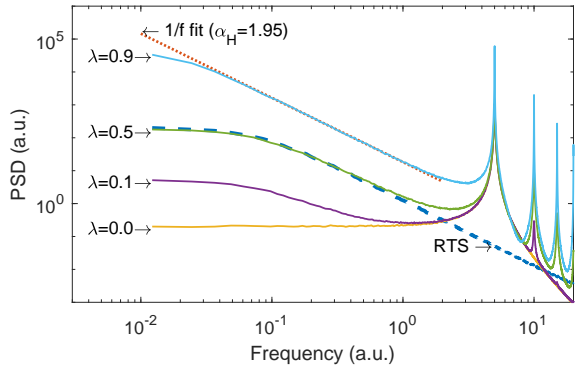}
	\caption{(Color online) Power spectral density (PSD) of the simulated waveforms in randomly driven bi-harmonic oscillators for different asymmetry parameter values. Dashed line is a spectrum of adjusted RTS (switching rates $r_{12}=1/3.5$ and $r_{21}=1/4.5$ with an amplitude of 5), which closely follows the spectrum of the $\lambda=0.5$ up to $f=1$. Dotted line is a $1/f^{\alpha_H}$ fit of the $\lambda=0.9$ spectrum in the range 0.1-1.0, producing Hooge coefficient $\alpha_H=1.95$.}
	\label{fig:biharmonicspectra}
	\end{figure}
For this demonstration we used $\tau_0=0.2$ and $\alpha=0.01$. FIG.~\ref{fig:biharmonicspectra} shows spectra of the simulated waveforms for different values of the asymmetry parameter. $\lambda=0$ corresponds to harmonic oscillations, and results in a single resonant peak with a low and flat LF tail. An expected result is the growing share of resonance harmonics with growing asymmetry. However, even small asymmetry also yields significant LF noise increase. 

The LF noise profile initially resembles random telegraph signal (RTS) spectra (with a frequency "knee"), making it a challenge to distinguish the discussed effect from random telegraph noise in frequency domain measurements. On the other hand, at high asymmetry it becomes close to 1/f noise with a Hooge coefficient $\alpha_H$ approaching 2. 

	\begin{figure}[!t]
	\centering
	\includegraphics[width=8.6 cm]{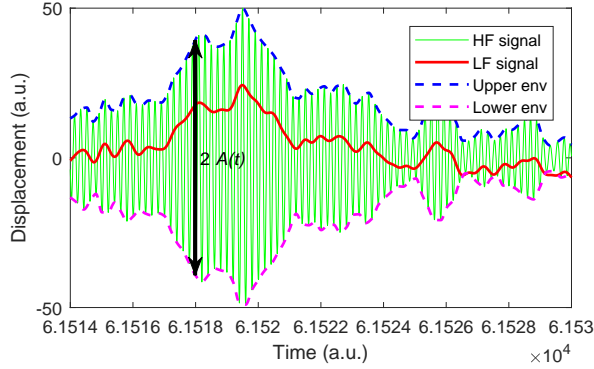}
	\caption{(Color online) A small section of the LF ($0<f<2$) and HF ($3<f<7$) signals extracted from the $\lambda=0.5$ waveform. Visual inspection confirms expected correlation between the LF signal and HF instantaneous amplitude $A(t)$. The latter is retrieved by finding upper and lower envelopes of the HF waveform.}
	\label{fig:envelopes}
	\end{figure}
This effect of LF noise arising from oscillation asymmetry originates from  the fact that the displacement averaged over a single period continues to vary significantly as the oscillation amplitude fluctuates (e.g. due to finite temperature). In order to illustrate this concept, the LF ($0<f<2$) and high frequency (HF, $3<f<7$) signals were extracted form the simulated waveform using zero phase digital filters \cite{Zerophasedigitalfiltering}.  The correlation between the LF signal and an instantaneous amplitude of HF oscillations is clearly visible in FIG.~\ref{fig:envelopes}. The amplitude evolution $A(t)$ can be found as a half of the difference between upper and lower envelopes of the HF waveform, as marked on the graph. 

	\begin{figure}[!t]
	\centering
	\includegraphics[width=8.6 cm]{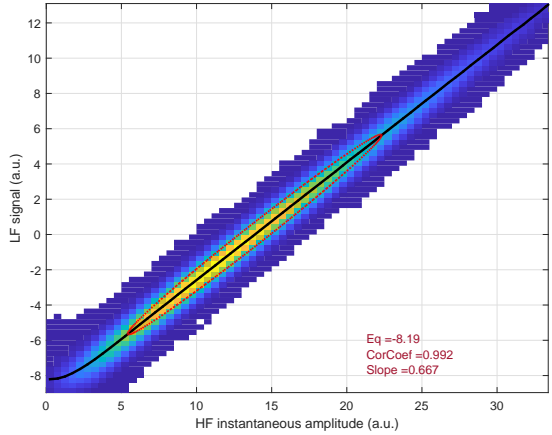}
	\caption{(Color online) 2-D histogram of LF signal ($f<2$) vs. the HF ($3<f<7$) instantaneous amplitude for a randomly driven bi-harmonic oscillator at asymmetry $\lambda=0.5$, showing a correlation coefficient equal to 0.992. Dotted (red) shape is a 0.5-confidence ellipse. Solid (black) line shows the regression curve, which is linear with a slope of 0.667 ($\approx 4 \lambda/\pi=0.637$). Small deviation from linearity at the lowest amplitudes ($A \lesssim 3$) is caused by limited accuracy of amplitude extraction. The equilibrium signal (characterized by Eq value), corresponding to the lowest amplitudes, is shifted from 0 due to simulated AC coupling. }
	\label{fig:biharmonichistogram}
	\end{figure}
FIG.~\ref{fig:biharmonichistogram} presents a two-dimensional histogram of LF signal vs. HF instantaneous amplitude. These variables are strongly correlated and the regression (defined as a LF signal mean value for each amplitude bin) is linear. Simple averaging of a single period shows that the slope of this regression line is expected to be equal $4 \lambda/\pi$, which is in good agreement with simulated data. 

\paragraph{Skew normal distribution oscillator model -}
The bi-harmonic oscillator model has attractive features for understanding asymmetric oscillations, as resonance frequency and full width at half maximum (FWHM) are independent of excitation amplitude. However, its energy profile is not analytic at the equilibrium point, and it is unlikely to be fully realized in experiments or devices. Actual asymmetric oscillations may reveal very different LF noise spectra. To gain  a stronger intuition about expected effects, other models have to be investigated. In particular, in a vertically flipped skew normal distribution (SND) function, asymmetry is controlled with a single real parameter $\lambda_s$: 
    \begin{equation}\label{eq:SND}
       E_{\lambda_s} (x) \sim - (1+\operatorname{erf}(\lambda_s x)) \exp(-x^2) \quad .
    \end{equation}

	\begin{figure}[!t]
	\captionsetup[subfigure]{labelformat=empty,farskip=4pt,captionskip=-12pt}
	\centering
	\subfloat[]{
	\includegraphics[width=8.6 cm]{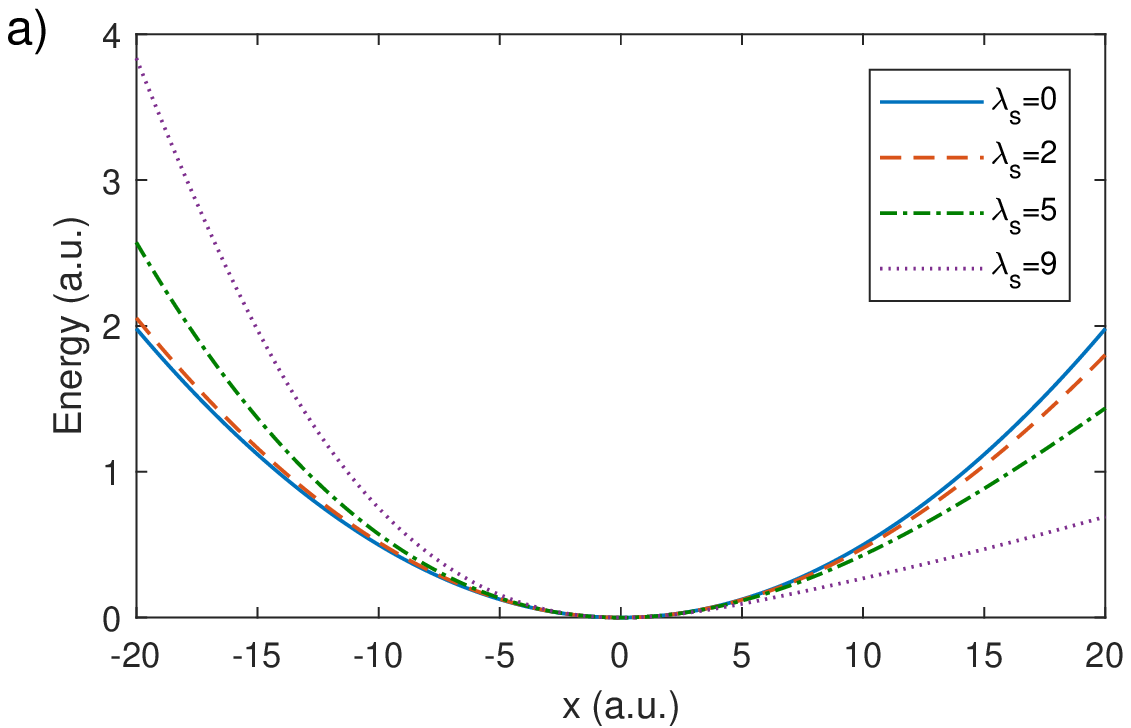}
	}
	\qquad	
	\subfloat[]{
	\includegraphics[width=8.6 cm]{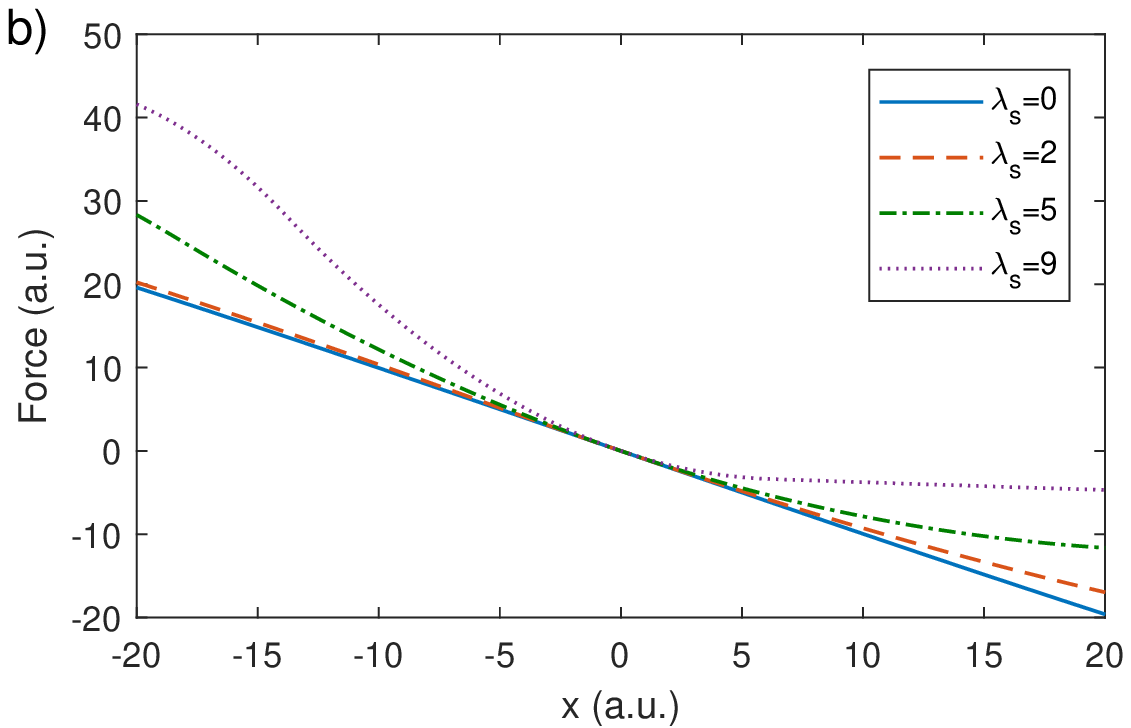}
	}
	\caption{(Color online) (a) Energy and (b) returning force of SND-based oscillator model for different values of the asymmetry parameter $\lambda_s$.}
	\label{fig:SNDenforce}
	\end{figure}
This function has been shifted and scaled horizontally and vertically to produce a force profile $F_{\lambda_s} (x)=-\partial E_{\lambda_s} (x)/\partial x$ with the slope equal $-1$ at $x=0$ for any $\lambda_s$, as illustrated in FIG.~\ref{fig:SNDenforce}. $\lambda_s=0$ corresponds to symmetric function derived from the normal distribution. 
The Langevin equation can be now written as 
    \begin{equation}\label{eq:LangevinSND}
	  \ddot{x} + 2 \beta_0 \dot{x} - (\omega_0)^2 F_{\lambda_s} (x) = \eta (t) \sqrt{\beta_0 \omega_0}	\quad ,
    \end{equation}
where $\beta_0=\alpha \omega_0$ ($\alpha$ is a damping constant). For simplicity, $\beta_0$ was assumed to remain independent of $x$. For $x \ll 1$, we have $F_{\lambda_s} (x) \approx -x$ , and   the model becomes a damped harmonic oscillator with undamped resonant frequency $\omega_0/{2 \pi}$.  

	\begin{figure}[!t]
	\centering
	\includegraphics[width=3.45in]{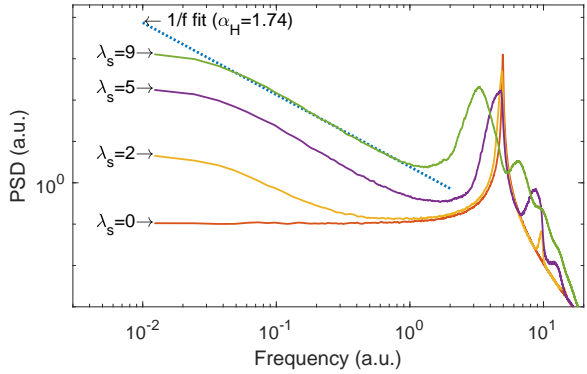}
	\caption{(Color online) PSD of the simulated waveforms in randomly driven SND-based oscillators for different asymmetry parameter values. Dotted line is a $1/f^{\alpha_H}$ fit of the $\lambda_s=9$ spectrum in the range 0.1-1.0, producing Hooge coefficient $\alpha_H=1.74$.}
	\label{fig:SNDspectra}
	\end{figure}
In the numerical integration of Eq. \ref{eq:LangevinSND}, a piece-wise, polynomial approximation of SND \cite{AshourApproximateskewnormal2010} was used, with $\tau_0=0.2$ and $\alpha=0.005$. The results are presented in FIG.~\ref{fig:SNDspectra}. The resonance line broadening and shift are expected effects of non-linearity. However, the presence of the LF tails is again a sole effect of the oscillation asymmetry. At strong asymmetry levels they also approach 1/f profile, now with the Hooge parameter of $\sim 1.74$ for $\lambda_s=9$. 

	\begin{figure}[!t]
	\centering
	\includegraphics[width=8.6 cm]{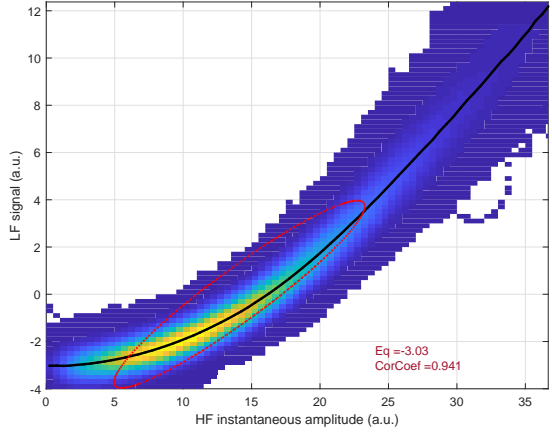}
	\caption{(Color online) 2-D histogram of LF noise vs. the HF instantaneous amplitude for a randomly driven SND-based oscillator at asymmetry $\lambda_s=5$, showing a correlation coefficient equal to 0.941.  Dotted (red) shape is a 0.5-confidence ellipse. Solid (black) line shows the regression curve. }
	\label{fig:SNDhistogram}
	\end{figure}
FIG.~\ref{fig:SNDhistogram} shows a 2-D histogram of LF signal and HF instantaneous amplitude for the asymmetric SND-based oscillator. The strong correlation between these signal components appears to be an ultimate test of noise generated by asymmetric oscillations. The regression line is non-linear here, indicating the non-linearity of the system. A comprehensive report of asymmetric oscillation modeling will be published separately \citep{Pipathanapoompronnthermallyexcitednoise}. 

\paragraph{Experimental evidence -}
The asymmetric oscillation noise has been first proposed as a source of LF noise in magnetic readers \cite{StankiewiczLowfrequencythermal2016}, in order to explain the unexpected M-shape noise dependence on external magnetic field. The latter was not compatible with any reader noise generation mechanisms reported before \cite{LeiReviewNoiseSources2011}. However, that experiment constituted merely an indirect argument to support the hypothesis. 

Magnetic oscillations correspond to a magnetization vector precession around the effective magnetic field. However, the angles describing this process are not independent, and can be reduced to 1-D problem \cite{StankiewiczDampingconstantestimation2015}, which justifies the analogy to the discussed models.  

	\begin{figure}[!t]
	\centering
	\includegraphics[width=8.6 cm]{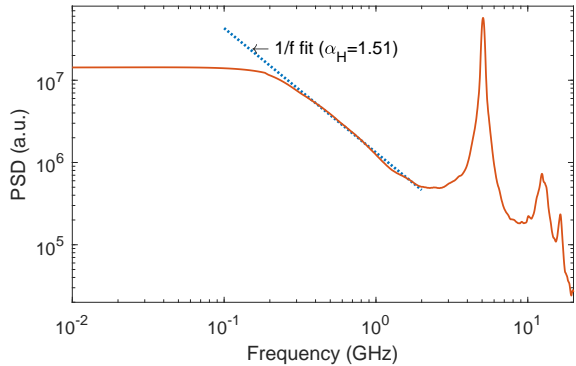}
	\caption{(Color online) PSD of the noise waveform from a noisy magnetic reader. Dotted line is a $1/f^{\alpha_H}$ fit in the range 0.3-1.0, producing Hooge coefficient $\alpha_H=1.51$.}
	\label{fig:expspectra}
	\end{figure}
	\begin{figure}[!t]
	\centering
	\includegraphics[width=8.6 cm]{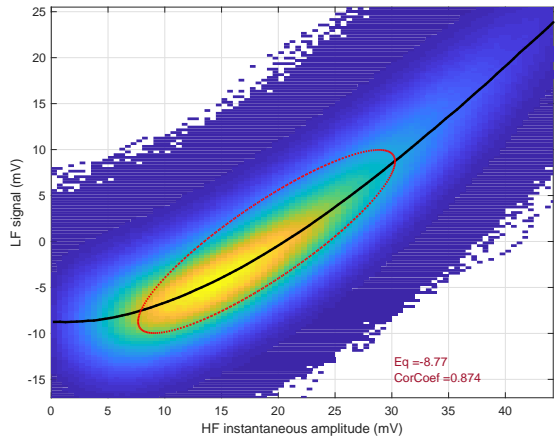}
	\caption{(Color online) 2-D histogram of LF signal ($f<2$ GHz) vs. the HF ($3$ GHz $<f<7$ GHz)  instantaneous amplitude derived from the magnetic reader noise waveform. The correlation coefficient is equal to 0.874. Dotted (red) curve is a 0.5-confidence ellipse. Solid (black) line shows the regression curve, which is very similar to the curve simulated under the SND-based oscillator model. }
	\label{fig:exphistogram}
	\end{figure}
The actual reader noise waveforms have been collected using a 90000-series Keysight oscilloscope, connected to the external output of an ISI FMR tester \cite{FMRA2008B}. The effective bandwidth of this setup was 18 GHz. FIG.~\ref{fig:expspectra} shows a spectrum of the selected noisy head. A prominent RTS-like LF noise profile is clearly visible here. However, no trace of any RTS process could be detected in the high bandwidth waveform. At the same time, a strong correlation between LF signal and HF instantaneous amplitude (FIG.~\ref{fig:exphistogram}) proves that the LF noise originates in thermally driven fluctuations of the asymmetric oscillation amplitude. 

It turns out that the phenomenon is present in all magnetic readers, but with different intensity \citep{StankiewiczReaderNoisedue2018}. Possible sources of the asymmetric output are currently under theoretical and experimental investigation. They include anisotropy of the magnetic structures \citep{Pipathanapoompronnthermallyexcitednoise} and non-linearity of the transfer function \cite{PauselliMagneticNoiseSpinWave2017}. Other factors are also analyzed, like non-linearity due to three-dimensional magnetization distributions, and  modal structure of thermal excitations \cite{PauselliLinearnonlineardynamics2017}. 

\paragraph{Conclusion -}
It appears that the low frequency noise excess is an immanent feature of all randomly stimulated asymmetric oscillators. The respective LF noise spectra may resemble either RTS or 1/f noise profiles, easily leading to confusion.  Detecting the effect is impossible in frequency domain data (i.e. based on power spectra), due to missing information about phase. However, high bandwidth signal in the time domain allows for identification of the effect by checking for a correlation between the LF portion of the signal and the HF amplitude fluctuations. This phenomenon has been experimentally confirmed in thermal noise of magnetoresistive sensors. 

More thorough analysis - analytic, numerical and experimental - is needed to fully understand all factors (e.g. statistics of the random stimulus, shape of the energy well, etc.) that may influence this kind of LF noise. It may be also expected that many more physical systems, other than magnetic readers, will show such noise when excitation is large enough to reveal oscillation asymmetry. 

\paragraph{Acknowledgment -}
The author is indebted to his Seagate colleagues: S. Stokes for inspiring this work,  T. Pipathanapoompron for help in SDE solver coding, K. Subramanian, T. Pokhil and V. Sapozhnikov for stimulating discussions.

\bibliography{StankiewiczPRLsubmit2018}
\end{document}